\documentclass[twocolumn,letter]{jpsj2}
\def\runauthor{\textsc{D.~Tahara} {\itshape et al}.}

\makeatletter
\def\@typeset{}
\makeatother
\def\journal #1#2#3#4{#1 {\bf #2} (#4) #3}
\def\PR{Phys.\ Rev.}
\def\PRB{Phys.\ Rev.\ B}
\def\PRL{Phys.\ Rev.\ Lett.}
\def\JPSJ{J.\ Phys.\ Soc.\ Jpn.}
\title{%
Antiferromagnetic Ising Model on Inverse Perovskite Lattice
}
\author{%
Daisuke \textsc{Tahara}, 
Yukitoshi \textsc{Motome}$^1$ and Masatoshi \textsc{Imada}$^1$
}
\inst{%
Department of Physics, University of Tokyo, 7-3-1 Hongo, Bunkyo-ku, Tokyo 113-0033\\
$^1$Department of Applied Physics, University of Tokyo, 7-3-1 Hongo, Bunkyo-ku, Tokyo 113-8656
}
\abst{%
We study thermodynamic properties of
an antiferromagnetic Ising model on the inverse perovskite lattice
by using Monte Carlo simulations.
The lattice structure is composed of corner-sharing octahedra and
contains three-dimensional geometrical frustration in terms of magnetic interactions.
The system with the nearest-neighbor interactions alone does not exhibit any phase transition, 
leading to a degenerate ground state with large residual entropy.
The degeneracy is lifted by an external magnetic field or by an anisotropy in the interactions.
Depending on the anisotropy, they stabilize either a 3D ferrimagnetic state 
or a partially-disordered antiferromagnetic (PDAF) state with a dimensionality reduction to 2D.
By the degeneracy-lifting perturbations,
all the transition temperatures of these different ordered states continuously grow from zero,
leaving an unusual zero-temperature critical point at the unperturbed point.
Such a zero-temperature multicriticality is not observed in other frustrated
structures such as face-centered cubic and pyrochlore.
The transition to the PDAF state is represented by either the
first- or second-order
boundaries separated by tricritical lines, whereas the PDAF phase
shows $1/3$ magnetization plateaus.
}
\kword{%
geometrical frustration, inverse perovskite, Ising model, 
residual entropy, magnetization plateau, partial disorder, 
dimensionality reduction, ferrimagnetism
}
\begin{document}
\maketitle
%
\begin{figure}[b]
\centering
\includegraphics[width=0.45\textwidth]{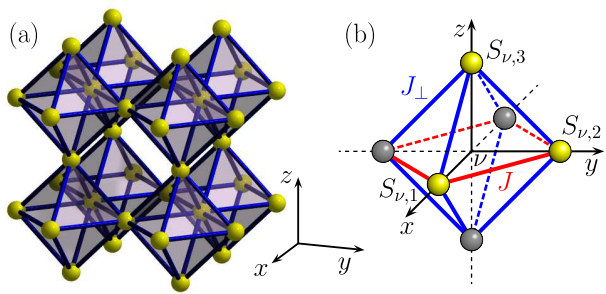}
\caption{%
(a) Inverse perovskite lattice structure. 
(b) A unit cell with three sublattice sites. 
$J$ and $J_\perp$ are the interactions in the Hamiltonian in eq.~(\ref{eq:H}).
}
\label{fig:01}
\end{figure}
\begin{figure}[b]
\centering
\includegraphics[width=0.45\textwidth]{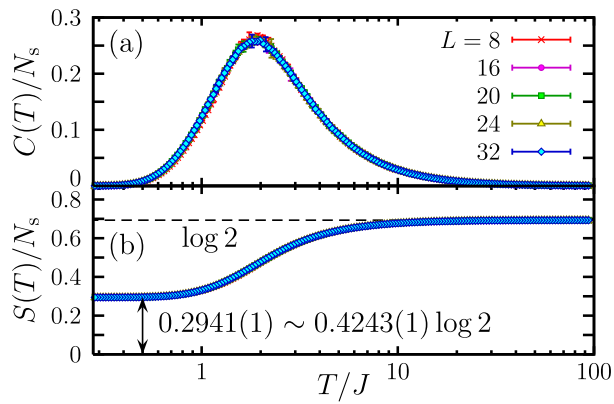}
\caption{%
Temperature dependences of (a) specific heat per site and (b) entropy per site at $J=J_\perp$ and $h=0$. 
The dashed line in (b) denotes the value in the high-temperature limit, $\log 2$.
}
\label{fig:02}
\end{figure}
Geometrical frustration is one of the long-standing issues in condensed matter physics.
The problem has been intensively investigated in antiferromagnetic (AF) spin systems.
Frustration, in general, suppresses the formation of simple-minded long-range orders (LRO) 
and opens the possibility of interesting phenomena
such as a degenerate ground-state manifold with an extensive residual entropy,
complicated ordering structures, and spin-liquid states\cite{Diep}.
\par
Even in one of the simplest models, 
namely the Ising model, 
frustration leads to intriguing properties
\cite{Diep,Liebmann}.
A prototypical example is the AF Ising model on a triangular lattice.
The system does not show any LRO down to zero temperature, 
and the ground state has macroscopic degeneracy\cite{Wannier,Houtappel,Husimi}.
Among other frustrated lattices, 
3D models have also attracted considerable attention
partly because certain real materials are available as their experimental counterparts.
For instance, 
the face-centered cubic (fcc) lattice has been studied
for the problem of binary alloys and
the pyrochlore lattice has been examined to describe
magnetite\cite{Anderson} and spin-ice compounds\cite{Ramirez}.
\par
In this paper, we discuss a new 3D frustrated Ising model, 
an inverse perovskite AF Ising model.
The inverse perovskite lattice consists of 
a simple cubic lattice of octahedra which share their corners,
as shown in Fig.~{\ref{fig:01}}(a).
The corner-sharing geometry reminds us of the well-studied pyrochlore lattice,
which consists of corner-sharing tetrahedra.
Alternatively, the lattice structure is viewed as an alternating stack of 
2D square lattices with different lattice constants
along the [001] direction (or the equivalent [100] or [010] direction),
while it is a stack of Kagom\'e lattices
along the $[111]$ direction (or the equivalent directions such as [$11\bar{1}$]). 
These stacks are similar to another well-studied lattice,
the fcc lattice, which is a stacking of 2D square (triangular) lattices 
along the $[001]$ ($[111]$) direction.
The inverse perovskite structure is found in several materials
such as the superconductor $\mathrm{Mg}\mathrm{C}\mathrm{Ni}_3$\cite{MgCNi3} 
and the negative thermal expansion material  
$\mathrm{Mn}_3(\mathrm{Cu}_{1-x}\mathrm{Ge}_x)\mathrm{N}$\cite{MnCuGeN}.
\par
%
In the present study,
we investigate effects of frustration on the thermodynamics
of the AF Ising model on the inverse perovskite lattice.
By employing a Monte Carlo (MC) method, 
we study phase transitions in this model
by tuning the degree of frustration and
obtain the phase diagram in a wide range of parameters.
In addition to some similar behaviors to the fcc and pyrochlore systems,
we find a peculiar critical point where all the transition temperatures
are suppressed down continuously to zero 
due to a characteristic structure of 
the degenerate ground-state manifold in the inverse perovskite system.
\par
Our model Hamiltonian is defined in the form
\begin{equation}
  \mathcal{H} = J \!\!\sum_{\langle i,j \rangle \in \parallel }\!\! S_i S_j
  + J_\perp \!\!\!\sum_{\langle i,j \rangle \in \perp }\!\! S_i S_j
  - h \sum_i S_i \, ,
  \label{eq:H}
\end{equation}
where $S_i = \pm 1$ denotes the Ising variable and 
$h$ is an external magnetic field.
In the inverse perovskite structure, three sites are present in a unit cell,
as shown in Fig.~\ref{fig:01}(b), and $S_i$ in eq.~(\ref{eq:H})
symbolically represents $S_{\nu,\alpha}$, where $\nu$ denotes the unit cell and
$\alpha = 1, 2$, or $3$ is the sublattice index.
The summations of the interaction terms are taken over
the nearest-neighbor sites
on the inverse perovskite lattice in Fig.~\ref{fig:01}(a), and
an anisotropy is introduced in the interactions $J$ and $J_\perp$,
as shown in Fig.~\ref{fig:01}(b):
$J$ is for pairs of $S_{\nu,1}$ and $S_{\nu',2}$ spins within the same $xy$ plane, 
and $J_\perp$ is for pairs including $S_{\nu,3}$ spins. 
We consider both cases of $J > J_\perp > 0$ and $J_\perp > J > 0$.
\par
We study thermodynamic properties of the model in eq.~(\ref{eq:H})
by using MC simulations.
To avoid the critical slowing down in the frustrated system, 
we employ the exchange MC method\cite{ExMC} 
by an exchange sampling taken after every 5 standard MC steps. 
Typically, we perform $10^5$--$10^7$ exchanges 
in total for the system size $L^3$ ($N_{\text{s}}=3L^3$ spins) 
up to $L=64$ under the periodic boundary conditions.
We set the Boltzmann constant $k_{\text{B}}=1$.
\par
%
First, we study the isotropic case $J=J_\perp$ at the magnetic field $h=0$.
Figure~\ref{fig:02}(a) shows the temperature ($T$) dependence of the specific heat per spin
plotted in the logarithmic scale of $T$.
The specific heat is calculated by
$C(T) = (\langle \mathcal{H}^2\rangle - \langle \mathcal{H}\rangle^2)/T^2$,
where the bracket denotes the thermal average.
As shown in Fig.~\ref{fig:02}(a), the specific heat shows no singularity, 
except for a broad hump at $T/J \sim 2$ which is almost system-size independent. 
This indicates that the model does not exhibit any phase transition at finite $T$.
The entropy is calculated by 
$
  S(T) = E(T)/T	- \int_{0}^{1/T}\!\!
	E(T') \,d(1/T')
  + S(\infty)\, ,
$
where $E(T)=\langle \mathcal{H} \rangle$ is the internal energy 
and $S(\infty)=N_{\text{s}}\log 2$ is the entropy in the high-$T$ limit. 
The result is plotted in Fig.~\ref{fig:02}(b). 
With lowering $T$, 
the entropy gradually decreases and approaches 
a finite value as $T \to 0$.
The residual entropy per spin is estimated as $0.2941(1) \sim 42$\% of $\log 2$
(statistical errors in the last digit are shown in the parentheses, hereafter). 
This result means that macroscopic degeneracy remains in the ground state. 
The absence of a phase transition and the large residual entropy are
found in other 3D frustrated AF Ising systems
such as pyrochlore\cite{Ramirez} and garnet\cite{spinice02}, 
and appears to be a common feature of corner-sharing-type frustrations.
\par
\begin{figure}[t]
\centering
\includegraphics[width=0.425\textwidth]{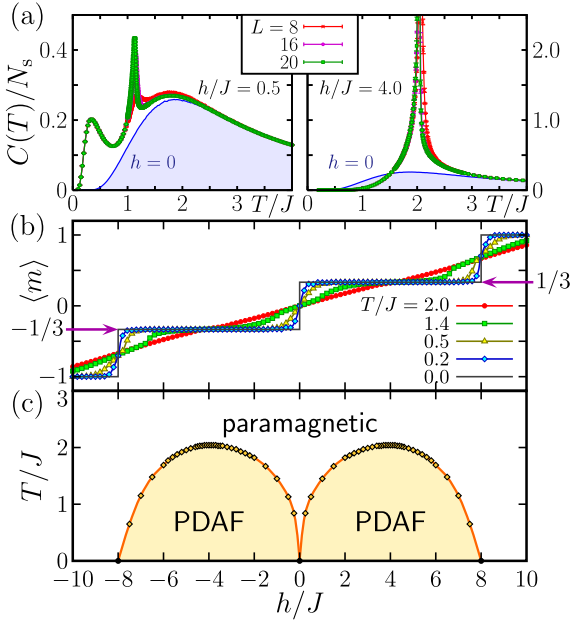}
\caption{%
(a) Temperature dependence of the specific heat at $h/J=0.5$ (left) and $4.0$ (right) 
for $J = J_\perp$.
The data at $h=0$ are shown for comparison.
(b) Magnetization process at several temperatures.
(c) Phase diagram at $J=J_\perp$. PDAF denotes the partially-disordered AF phase. 
The lines are guides for the eyes. 
}
\label{fig:03}
\end{figure}
\begin{figure}[t]
\centering
\includegraphics[width=0.425\textwidth]{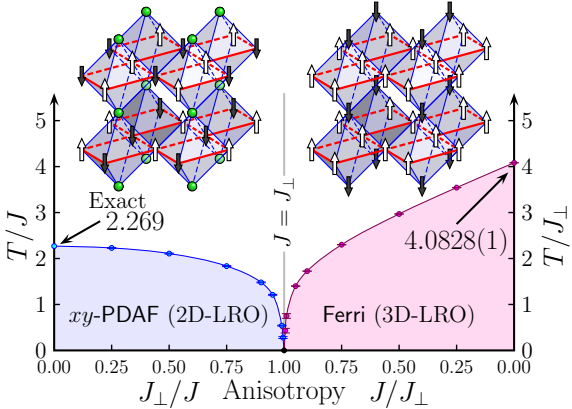}
\caption{%
Phase diagram for the anisotropy in the interaction $(J,J_\perp)$
at $h=0$. The lines are guides  for the eyes. 
Schematic pictures of $xy$-PDAF state (left) and 3D Ferri state (right) are also shown.
Open (filled) arrows denote up (down) spins, 
and balls represent disordered spins.
}
\label{fig:04}
\end{figure}

When a magnetic field is applied, the macroscopic degeneracy is lifted and
a phase transition takes place at finite $T$.
Figure~\ref{fig:03}(a) shows $T$ dependences of the specific heat.
The specific heat shows a sharp singularity for $|h|>0$, 
indicating a phase transition;
for instance, a sharp peak at $T/J \sim 1.1$ $(2.0)$ for $h/J = 0.5$ $(4.0)$
as shown in Fig.~\ref{fig:03}(a).
The small hump at $T/J \sim 0.3$ for $h/J=0.5$ comes from the 
disordered spins remaining in the low-$T$ ordered state (see below).
Below the transition temperature,
the magnetization process shows plateaus at $\langle m \rangle = \pm 1/3$ 
as shown in Fig.~\ref{fig:03}(b), 
where $m = \sum_{\nu,i} S_{\nu,i} / N_{\text{s}}$.
The plateau state shows a partially-disordered AF (PDAF) ordering,
which consists of the stacking of AF-ordered square lattices
and disordered spins sandwiched between AF layers,
as shown schematically in the left inset of Fig.~\ref{fig:04}.
This dimensionality reduction, i.e., 2D LRO in a 3D system, yields
$2^L$-fold degeneracy due to the twofold degeneracy
of the AF ordering pattern in each layer, 
leading to $(3\times 2^L -3)$-fold degeneracy in total. 
The factor $3$ comes from the stacking directions 
$x$, $y$, and $z$, 
and $-3$ subtracts the double counting.
The schematic picture in Fig.~\ref{fig:04} shows 
one of these degenerate states.
The magnetic ordering and degeneracy are confirmed by calculating 
the 2D AF order parameter in each $xy$, $yz$, and $zx$ layer.
We summarize the critical temperature $T_{\text{c}}$
determined from these magnetic properties
in the phase diagram in Fig.~\ref{fig:03}(c).
$T_{\text{c}}$ is finite at $0 < |h|/J < 8$, 
showing its maxima $T_{\rm c}/J \sim 2$ at $h/J \sim \pm 4$. 
The transition is of first order, confirmed by a detailed analysis of 
the probability distribution of the internal energy (not shown). 
\par
A similar dimensionality reduction has also been found in 
the ground state of the fcc AF Ising model\cite{Danielian}.
In addition, similar magnetization plateaus have also been found in 
both the fcc\cite{Lebowitz,Kammerer} and pyrochlore models\cite{Liebmann}.
In these systems, however, the plateau states appear 
above some finite critical value of the applied magnetic field.
A peculiar feature of our inverse perovskite model is that
the critical field is zero
and the plateau phases for the positive and negative values of $h$ touch with each other 
at $h=0$ with the complete suppression of $T_{\rm c}$,
as shown in Fig.~\ref{fig:03}(c).
\par
%
The macroscopic degeneracy in the case of $J=J_\perp$ and $h=0$
is also lifted by introducing an anisotropy in the interaction, namely, $J \neq J_\perp$.
First, let us consider the case of $J>J_\perp$.
In the limit of $J_\perp/J \to 0$, the system is decomposed into 
independent 2D square lattices and disconnected sites.
Therefore, the system undergoes the same phase transition as that in the 2D Ising model
at $T_{\text{c}}/J = 2/\sinh^{-1} 1 \simeq 2.269$\cite{Onsager} and
the ordered phase is the PDAF state in which the AF orderings occur independently
in the $xy$ square lattices.
This $xy$-PDAF state is stabilized at low $T$ 
in the entire region of $J>J_\perp$.
The order parameter is the staggered magnetization of each $xy$ plane;
$
  m_{xy} = \sum'_{\nu} \bigl( S_{\nu,1} - S_{\nu,2} \bigr) / 2L^2\, ,
$
where the summation is taken over the spins within one $xy$ layer.
The transition is of second order, and 
the critical temperature $T_{\text{c}}$ is determined by 
the Binder parameter\cite{jjj}
$g_{xy} = 1 - \langle m_{xy}^4 \rangle / 3 \langle m_{xy}^2\rangle^2$.  
The results are plotted in the left half of Fig.~\ref{fig:04}. 
$T_{\text{c}}$ decreases as $J_\perp/J$ increases and goes continuously to zero
when $J_\perp/J \to 1$.
We also examined the universality class of this $xy$-PDAF transition
by the finite-size scaling (FSS) analysis;
$\langle m_{xy}^2 \rangle = L^{-2\beta/\nu} f(tL^{1/\nu})$, 
where $f$ is a scaling function and $t=T/T_{\text{c}}-1$ is the reduced temperature. 
We confirmed that the transition belongs to 
the 2D Ising universality with the exponents $\beta=1/8$ and $\nu=1$\cite{2DIsingExp}. 
\par
%
\begin{figure}[t]
\centering
\includegraphics[width=0.425\textwidth]{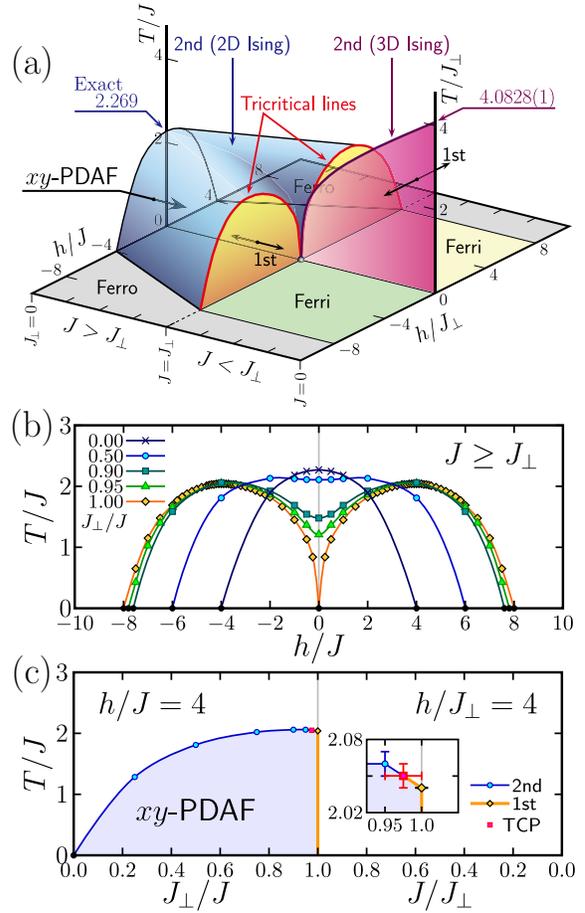}
\caption{%
(a) Phase diagram of the inverse perovskite AF Ising model in the $T$-$h$-$(J,J_\perp)$ space.
(b) $h$-$T$ phase diagrams for several values of $J_\perp/J$ for $J \geq J_\perp$.
(c) $(J,J_\perp)$-$T$ phase diagram at $h = 4 \max(J,J_\perp)$. 
The lines are guides  for the eyes, and TCP denotes the tricritical point.
}
\label{fig:05}
\end{figure}
%
\par
On the other hand, on the opposite side, i.e., $J<J_\perp$,
we find a phase transition to the ferrimagnetic (Ferri) 3D-LRO state, 
as shown schematically in the right inset of Fig.~\ref{fig:04}. 
The order parameter is defined by 
$
  m_{f} = \sum_{\nu} \bigl( S_{\nu,1} + S_{\nu,2} - S_{\nu,3} \bigr) / 3L^3 \, .
$
The Ferri ground state has twofold degeneracy, $m_{f} = \pm 1/3$
corresponding to the spin configurations 
$(S_{\nu,1},S_{\nu,2},S_{\nu,3})=(1,1,-1)$ and $(-1,-1,1)$, respectively. 
In this case also, the transition is of the second order, and 
the values of $T_{\text{c}}$ obtained by the Binder analysis are 
plotted in the right half of Fig.~\ref{fig:04}.
$T_{\text{c}}$ decreases as $J/J_\perp$ increases and
goes continuously to zero when $J/J_\perp \to 1$,
as in the case of $J>J_\perp$.
We also confirmed that the transition belongs to the 3D Ising universality class
by the FSS analysis.
For instance, in the limit of $J/J_\perp \to 0$, 
we obtained $T_{\text{c}}/J_\perp = 4.0828(1)$ and
the critical exponents $\beta=0.32(1)$ and $\nu=0.625(7)$, 
which are consistent with the 3D Ising exponents, 
$\beta=0.325$ and $\nu=0.63$\cite{3DIsingExp}, within the statistical errors. 
\par
The effect of anisotropy $J/J_\perp$ at $h=0$ is summarized as follows: 
The anisotropy lifts the macroscopic degeneracy and induces
the phase transition at finite $T$;
the $xy$-PDAF state appears for $J>J_\perp$ and the 3D Ferri state for $J<J_\perp$.
The two different ordered phases touch with each other 
at the isotropic point $J=J_\perp$.
Here, $T_{\rm c}$ goes continuously to zero on both the sides.
The universality class of the transition is different 
between the cases of $J>J_\perp$ and $J<J_\perp$;
the 2D Ising one in the former 
due to the dimensionality reduction
and the 3D Ising one in the latter.
These properties are contrasted to those of the fcc system. 
In the isotropic fcc model, a first-order phase transition takes place 
at finite $T$, namely, 
there is no suppression of $T_{\text{c}}$\cite{Danielian, Phani}.
It was also proposed that the critical exponents 
change continuously\cite{2-1-4Ising} 
when a similar anisotropy in the interactions is introduced.
\par
On the basis of the above analyses obtained from applying magnetic field $h$ or 
from introducing the anisotropy $J \neq J_\perp$,
the peculiarity of the isotropic model with $J=J_\perp$ and $h=0$ is revealed:
The model has a critical point at $T=0$ in the sense that
the transition temperatures are suppressed
down continuously to zero 
in both the cases of $|h| \to 0$ and $J/J_\perp \to 1$.
\par
To further clarify the behavior around the critical point,
we investigate the model by changing both the 
$h$ and $J/J_\perp$ values. 
The phase diagram in a wide range of parameters is shown in Fig.~\ref{fig:05}(a).
When we decrease $J_\perp/J$ from 1,
two PDAF phases in the case of $J=J_\perp$ in Fig.~\ref{fig:03}(c) 
gradually merge into one,
as shown in Fig.~\ref{fig:05}(b).
On the other hand, in the region of $J<J_\perp$,
the magnetic field is a symmetry-breaking field of the Ferri ordering at $h=0$, 
and therefore,
the finite-$T$ phase transition disappears at $h\neq 0$.
Hence, at finite $h$, the PDAF ordered phase 
suddenly terminates at $J=J_\perp$.
This behavior is studied in detail by changing $J/J_\perp$.
For instance, Fig.~\ref{fig:05}(c) shows the result at $h=4 \max(J,J_\perp)$.
Since the phase transition at the isotropic case $J=J_\perp$ is
of the first order as discussed earlier 
and since no transition exists at finite $T$ for $J<J_\perp$,
we have a discontinuous phase boundary at $J=J_\perp$ below $T_{\text{c}}$.
This discontinuous boundary slightly extends to $J>J_\perp$ 
and terminates at a tricritical point 
beyond which the transition becomes continuous,
as shown in Fig.~\ref{fig:05}(c). 
The tricritical behavior appears for $0 < |h|/J <8$,
resulting in the tricritical lines which start from $(J_\perp/J, h/J, T/J) = (1, 0, 0)$
and end at $(1, \pm 8, 0)$, as shown in Fig.~\ref{fig:05}(a).
Note that the tricritical lines are not in the $J=J_\perp$ plane,
but in a very narrow region, $0.95 < J_\perp/J \leq 1$.
\par
%
All the features above are shown in Fig.~\ref{fig:05}(a)
with the ground-state phase diagram which is easily obtained
by minimizing the energy of one octahedron.
The figure includes the PDAF phase for $J>J_\perp$ (blue region), 
the discontinuous surface in the $h=0$ plane for $J<J_\perp$ (purple plane),
and the tricritical lines (red curves) 
surrounding the discontinuous phase boundaries at the edges of 
the PDAF state (orange surfaces).
This global phase diagram clearly demonstrates the peculiarity of
the critical point at $J=J_\perp$ and $h=0$,
where all the critical temperatures, for both first- and second-order transitions, 
are suppressed down continuously to zero.
\par
The absence of LRO at the critical point is understood as follows.
When we start from a PDAF ordered state,
we can generate an anti-phase domain in the AF-ordered square lattice
without any energy cost
because of the partial disorder.
This leads to the destruction of PDAF ordering and 
to the formation of a degenerate ground-state manifold.
The situation is similar when we start from the Ferri 3D-LRO state.
These anti-phase domains cost a finite energy when we introduce 
the magnetic field $h$ and/or anisotropy $J/J_\perp$
and hence finite-$T$ phase transitions immediately emerge around the critical point.
\par
In summary, by using the exchange Monte Carlo method, 
the phase diagram of the AF Ising model on the inverse perovskite lattice has been clarified 
in the parameter space of the uniform magnetic field and the anisotropy in the interaction. 
It has revealed an unusual character of the $T=0$ critical point,
which drags the surrounding critical lines and surfaces at $T\ne 0$ toward absolute zero temperature.
Neither fcc nor pyrochlore AF Ising model exhibits such a behavior.
\par
In the fcc system at the isotropic point, there is no suppression of $T_{\text{c}}$
and a first-order transition to the 2D-LRO state takes place\cite{Danielian, Phani}.
In the isotropic pyrochlore system,
although the ground state has macroscopic degeneracy as that in the present system,
the degeneracy remains until a finite magnetic field is applied ($|h| < 2J$)\cite{Liebmann}.
\par
The pyrochlore system has attracted interest
due to the degenerate ground state manifold, and 
the effects of the Heisenberg-type interaction 
and quantum fluctuations have been examined intensively
in the search for an ``order from disorder'' phenomenon or a spin-liquid state.
Since the degenerate ground state in the inverse perovskite model
is at a more unusual critical state with competing orders,
it would be more interesting and worthwhile
to study such effects in the inverse perovskite system.
In particular, quantum fluctuations may easily destroy the orderings 
barely stabilized around the $T=0$ classical critical point, 
which may open possibilities for quantum spin-liquid phases or exotic quantum orders. 
\par
%
One of the authors (D.T.) thanks T.~Misawa for useful discussions.
This work was supported by Grants-in-Aid for Scientific Research on Priority Areas under the grant numbers 
17071003, 
16076212, 
17740244, and 
16GS0219
from the Ministry of Education, Culture, Sports, Science and Technology. 
A part of our computation has been done using the facilities of the Supercomputer Center, 
Institute for Solid State Physics, University of Tokyo.

%
%
\end{document}